\documentclass
[prl,amsmath,amssymb,reprint,superscriptaddress,showpacs]
{revtex4-1}
\usepackage{color}
\usepackage{eso-pic} 
\usepackage{amssymb}
\usepackage{feynmp}
\usepackage{comment}
\usepackage{bm}
\usepackage{slashbox}
\usepackage{multicols} 
\usepackage{graphicx}

\newif\ifpdf  
 \ifx\pdfoutput\undefined
 \pdffalse                  
\else 
 \pdfoutput=1               
 \pdftrue 
\fi  
     
\ifpdf    
 \usepackage{url,boxedminipage}
 \usepackage{graphicx,thumbpdf}
 \definecolor{rltred}{rgb}{0.75,0,0}
 \definecolor{rltgreen}{rgb}{0,0.5,0}
 \definecolor{rltblue}{rgb}{0,0,0.75}
 \usepackage[colorlinks%
  ,bookmarks%
  ,urlcolor=rltblue%
  ,filecolor=rltgreen%
  ,linkcolor=rltred%
  ,pdftitle={pdftitLe}%
  ,pdfsubject={pdfsubject}%
  ,pdfkeywords={pdfkeywords}%
  ,pdfauthor={Joe User <hirokazu@fnal.gov>}%
  ,pdfpagemode={UseOutlines}%
  ,bookmarksopen=true%
  ,bookmarksnumbered=true%
  ,pdfstartview={Fit}%
  ]{hyperref}
 \DeclareGraphicsExtensions{.pdf,.jpg,.jpeg}            
\else
 \usepackage{graphicx}
 \DeclareGraphicsExtensions{.eps,.ps,.eps.gz,.ps.gz}    

\fi 

\def\subdef#1{\gdef\globalColor##1{##1}}
\def\newColor #1 {\expandafter\def\csname #1\endcsname##1{##1}
   \expandafter\def\csname text#1\endcsname{\subdef{#1}
    }}%
\newColor GreenYellow     
\newColor Yellow          
\newColor Goldenrod       
\newColor Dandelion       
\newColor Apricot         
\newColor Peach           
\newColor Melon           
\newColor YellowOrange    
\newColor Orange          
\newColor BurntOrange     
\newColor Bittersweet     
\newColor RedOrange       
\newColor Mahogany        
\newColor Maroon          
\newColor BrickRed        
\newColor Red             
\newColor OrangeRed       
\newColor RubineRed       
\newColor WildStrawberry  
\newColor Salmon          
\newColor CarnationPink   
\newColor Magenta         
\newColor VioletRed       
\newColor Rhodamine       
\newColor Mulberry        
\newColor RedViolet       
\newColor Fuchsia         
\newColor Lavender        
\newColor Thistle         
\newColor Orchid          
\newColor DarkOrchid      
\newColor Purple          
\newColor Plum            
\newColor Violet          
\newColor RoyalPurple     
\newColor BlueViolet      
\newColor Periwinkle      
\newColor CadetBlue       
\newColor CornflowerBlue  
\newColor MidnightBlue    
\newColor NavyBlue        
\newColor RoyalBlue       
\newColor Blue            
\newColor Cerulean        
\newColor Cyan            
\newColor ProcessBlue     
\newColor SkyBlue         
\newColor Turquoise       
\newColor TealBlue        
\newColor Aquamarine      
\newColor BlueGreen       
\newColor Emerald         
\newColor JungleGreen     
\newColor SeaGreen        
\newColor Green           
\newColor ForestGreen     
\newColor PineGreen       
\newColor LimeGreen       
\newColor YellowGreen     
\newColor SpringGreen     
\newColor OliveGreen      
\newColor RawSienna       
\newColor Sepia           
\newColor Brown           
\newColor Tan             
\newColor Gray            
\newColor Black           
\newColor White           

\subdef{Black}

\def \blankline {$\mbox{}$\\}

\def \mc {\multicolumn}

\newcommand{\type}[1]{{\tt #1}}

\def \bm#1{ \mbox{\boldmath $#1$} }

\def \bmath {\begin{displaymath}}
\def \emath {\end  {displaymath}}
\def \beq   {\begin{equation}}
\def \eeq   {\end  {equation}}
\def \beqa   {\begin{eqnarray}}
\def \eeqa   {\end  {eqnarray}}
\def \bitem {\begin{itemize}}
\def \eitem {\end  {itemize}}
\def \benum {\begin{enumerate}}
\def \eenum {\end  {enumerate}}
\def \bcenter {\begin{center}}
\def \ecenter {\end  {center}}
\def \CommentOut#1 {\if0 #1 \fi}
\def \bfig { \begin{figure} }
\def \efig { \end  {figure} }
\def \barray {\begin{array}}
\def \earray {\end  {array}}

\newlength{\pushupfigure}
\def \epsfin_v1#1#2{
	\vspace{\pushupfigure}
	\center
	\leavevmode
	\epsfxsize=#1
	\epsffile[20 143 575.75 698.75]{#2}
}
\newlength{\figsize     } 
\newlength{\oneplot     } 
\newlength{\plotsize    } 
\newlength{\twoplots    } 
\newlength{\twoplotsnote} 
\newlength{\oneplotevd  } 
\newlength{\twoplotsevd } 

\def \OneGSize#1 {
\includegraphics[width=\oneplot]{#1}
}

\def \TwoGSize#1 {
\includegraphics[width=\twoplots]{#1}
}

\def \TwoGSizeNote#1 {
\includegraphics[width=\twoplotsnote]{#1}
}

\makeatletter
  \def\flushpage{
    \ifodd\c@page
      \clearpage\blankline\clearpage
    \else
      \clearpage
    \fi
  }
\makeatother

\def \TwoC#1#2 {
\vspace{5mm}
\begin{multicols}{2}
\begin{minipage}[h]{0.92\linewidth}
#1
\vspace{-2cm}
\end{minipage}
\begin{minipage}[h]{0.92\linewidth}
#2
\vspace{-2cm}
\end{minipage}
\end{multicols}
}

\def \wrap#1{\mbox{$#1$}}

\def \to     {\wrap{\rightarrow}}

\def \isolmuon {$E_{T}^{\rm cone}$}
\def \misolmuon {E_{T}^{\rm cone}}
\def \isolmuon2 {$P_{T}^{\rm cone}$}
\def \misolmuon2 {P_{T}^{\rm cone}}

\def \mgev   {GeV/$c^{2}$}
\def \pgev   {GeV/$c$}

\def \et     {\wrap{E_{T}}}

\def \msp    {\:}

\def \jnewpagetitle#1	{\newpage  \begin{flushright} 
			 {\jnormalsize --- #1 ---}
			 \end{flushright}}
\def \newpagetitle#1	{\newpage  \begin{flushright} 
			 {\normalsize \sf --- #1 ---}
			 \end{flushright}}

\newlength{\plotwidth}
\newlength{\plotheight}
\def \type#1 {{\tt #1 }}

\def \BF     {\wrap{\cal B}}
\def \hmass  {\wrap{m_{h_f}}}
\def \Hmass  {\wrap{m_{H^\pm}}}
\def \nobs   {n}
\def \Vnobs  {\wrap{\bm{\nobs}}}
\def \E      {\wrap{\bm{E}}}

\def \Vntrue {\wrap{\bm{n}^*}}

\def \etsum    {\wrap{E^{\gamma_1}_T + E^{\gamma_2}_T}}

\def \pythia   {{\sc pythia}}
\def \geant    {{\sc geant}}
\def \mg       {{\sc MadGraph}}
\def \me       {{\sc MadEvent}}
\def \mgme     {\mg}
\def \mcfm     {{\sc mcfm}}

\def \DTPSysXSec      {0.83 events}

\def\mhMin {10}
\def\mhMax {105}
\def\mHMin {30}
\def\mHMax {300}



\renewcommand{\twoplotsnote}{0.50\textwidth}
\renewcommand{\twoplots    }{0.50\textwidth}

\def \fdir  {./figures}

\def \CL    {credibility}
\def \exampleMasses {a signal point having
$\hmass=75$~\mgev\ and $\Hmass=120$~\mgev}
\def \xsec     {cross section}
\def \xsechyph {cross-section}
\begin{document}
\title{
Search for a Low-Mass Neutral Higgs Boson with Suppressed Couplings 
to Fermions 
\\[0.2cm]
Using Events with Multiphoton Final States 
}

\affiliation{Institute of Physics, Academia Sinica, Taipei, Taiwan 11529, Republic of China}
\affiliation{Argonne National Laboratory, Argonne, Illinois 60439, USA}
\affiliation{University of Athens, 157 71 Athens, Greece}
\affiliation{Institut de Fisica d'Altes Energies, ICREA, Universitat Autonoma de Barcelona, E-08193, Bellaterra (Barcelona), Spain}
\affiliation{Baylor University, Waco, Texas 76798, USA}
\affiliation{Istituto Nazionale di Fisica Nucleare Bologna, \ensuremath{^{kk}}University of Bologna, I-40127 Bologna, Italy}
\affiliation{University of California, Davis, Davis, California 95616, USA}
\affiliation{University of California, Los Angeles, Los Angeles, California 90024, USA}
\affiliation{Instituto de Fisica de Cantabria, CSIC-University of Cantabria, 39005 Santander, Spain}
\affiliation{Carnegie Mellon University, Pittsburgh, Pennsylvania 15213, USA}
\affiliation{Enrico Fermi Institute, University of Chicago, Chicago, Illinois 60637, USA}
\affiliation{Comenius University, 842 48 Bratislava, Slovakia; Institute of Experimental Physics, 040 01 Kosice, Slovakia}
\affiliation{Joint Institute for Nuclear Research, RU-141980 Dubna, Russia}
\affiliation{Duke University, Durham, North Carolina 27708, USA}
\affiliation{Fermi National Accelerator Laboratory, Batavia, Illinois 60510, USA}
\affiliation{University of Florida, Gainesville, Florida 32611, USA}
\affiliation{Laboratori Nazionali di Frascati, Istituto Nazionale di Fisica Nucleare, I-00044 Frascati, Italy}
\affiliation{University of Geneva, CH-1211 Geneva 4, Switzerland}
\affiliation{Glasgow University, Glasgow G12 8QQ, United Kingdom}
\affiliation{Harvard University, Cambridge, Massachusetts 02138, USA}
\affiliation{Division of High Energy Physics, Department of Physics, University of Helsinki, FIN-00014, Helsinki, Finland; Helsinki Institute of Physics, FIN-00014, Helsinki, Finland}
\affiliation{University of Illinois, Urbana, Illinois 61801, USA}
\affiliation{The Johns Hopkins University, Baltimore, Maryland 21218, USA}
\affiliation{Institut f\"{u}r Experimentelle Kernphysik, Karlsruhe Institute of Technology, D-76131 Karlsruhe, Germany}
\affiliation{Center for High Energy Physics: Kyungpook National University, Daegu 702-701, Korea; Seoul National University, Seoul 151-742, Korea; Sungkyunkwan University, Suwon 440-746, Korea; Korea Institute of Science and Technology Information, Daejeon 305-806, Korea; Chonnam National University, Gwangju 500-757, Korea; Chonbuk National University, Jeonju 561-756, Korea; Ewha Womans University, Seoul, 120-750, Korea}
\affiliation{Ernest Orlando Lawrence Berkeley National Laboratory, Berkeley, California 94720, USA}
\affiliation{University of Liverpool, Liverpool L69 7ZE, United Kingdom}
\affiliation{University College London, London WC1E 6BT, United Kingdom}
\affiliation{Centro de Investigaciones Energeticas Medioambientales y Tecnologicas, E-28040 Madrid, Spain}
\affiliation{Massachusetts Institute of Technology, Cambridge, Massachusetts 02139, USA}
\affiliation{University of Michigan, Ann Arbor, Michigan 48109, USA}
\affiliation{Michigan State University, East Lansing, Michigan 48824, USA}
\affiliation{Institution for Theoretical and Experimental Physics, ITEP, Moscow 117259, Russia}
\affiliation{University of New Mexico, Albuquerque, New Mexico 87131, USA}
\affiliation{The Ohio State University, Columbus, Ohio 43210, USA}
\affiliation{Okayama University, Okayama 700-8530, Japan}
\affiliation{Osaka City University, Osaka 558-8585, Japan}
\affiliation{University of Oxford, Oxford OX1 3RH, United Kingdom}
\affiliation{Istituto Nazionale di Fisica Nucleare, Sezione di Padova, \ensuremath{^{ll}}University of Padova, I-35131 Padova, Italy}
\affiliation{University of Pennsylvania, Philadelphia, Pennsylvania 19104, USA}
\affiliation{Istituto Nazionale di Fisica Nucleare Pisa, \ensuremath{^{mm}}University of Pisa, \ensuremath{^{nn}}University of Siena, \ensuremath{^{oo}}Scuola Normale Superiore, I-56127 Pisa, Italy, \ensuremath{^{pp}}INFN Pavia, I-27100 Pavia, Italy, \ensuremath{^{qq}}University of Pavia, I-27100 Pavia, Italy}
\affiliation{University of Pittsburgh, Pittsburgh, Pennsylvania 15260, USA}
\affiliation{Purdue University, West Lafayette, Indiana 47907, USA}
\affiliation{University of Rochester, Rochester, New York 14627, USA}
\affiliation{The Rockefeller University, New York, New York 10065, USA}
\affiliation{Istituto Nazionale di Fisica Nucleare, Sezione di Roma 1, \ensuremath{^{rr}}Sapienza Universit\`{a} di Roma, I-00185 Roma, Italy}
\affiliation{Mitchell Institute for Fundamental Physics and Astronomy, Texas A\&M University, College Station, Texas 77843, USA}
\affiliation{Istituto Nazionale di Fisica Nucleare Trieste, \ensuremath{^{ss}}Gruppo Collegato di Udine, \ensuremath{^{tt}}University of Udine, I-33100 Udine, Italy, \ensuremath{^{uu}}University of Trieste, I-34127 Trieste, Italy}
\affiliation{University of Tsukuba, Tsukuba, Ibaraki 305, Japan}
\affiliation{Tufts University, Medford, Massachusetts 02155, USA}
\affiliation{Waseda University, Tokyo 169, Japan}
\affiliation{Wayne State University, Detroit, Michigan 48201, USA}
\affiliation{University of Wisconsin-Madison, Madison, Wisconsin 53706, USA}
\affiliation{Yale University, New Haven, Connecticut 06520, USA}

\author{T.~Aaltonen}
\affiliation{Division of High Energy Physics, Department of Physics, University of Helsinki, FIN-00014, Helsinki, Finland; Helsinki Institute of Physics, FIN-00014, Helsinki, Finland}
\author{S.~Amerio\ensuremath{^{ll}}}
\affiliation{Istituto Nazionale di Fisica Nucleare, Sezione di Padova, \ensuremath{^{ll}}University of Padova, I-35131 Padova, Italy}
\author{D.~Amidei}
\affiliation{University of Michigan, Ann Arbor, Michigan 48109, USA}
\author{A.~Anastassov\ensuremath{^{w}}}
\affiliation{Fermi National Accelerator Laboratory, Batavia, Illinois 60510, USA}
\author{A.~Annovi}
\affiliation{Laboratori Nazionali di Frascati, Istituto Nazionale di Fisica Nucleare, I-00044 Frascati, Italy}
\author{J.~Antos}
\affiliation{Comenius University, 842 48 Bratislava, Slovakia; Institute of Experimental Physics, 040 01 Kosice, Slovakia}
\author{G.~Apollinari}
\affiliation{Fermi National Accelerator Laboratory, Batavia, Illinois 60510, USA}
\author{J.A.~Appel}
\affiliation{Fermi National Accelerator Laboratory, Batavia, Illinois 60510, USA}
\author{T.~Arisawa}
\affiliation{Waseda University, Tokyo 169, Japan}
\author{A.~Artikov}
\affiliation{Joint Institute for Nuclear Research, RU-141980 Dubna, Russia}
% ...

\collaboration{CDF Collaboration}
\altaffiliation[With visitors from]{
\ensuremath{^{a}}University of British Columbia, Vancouver, BC V6T 1Z1, Canada,
\ensuremath{^{b}}Istituto Nazionale di Fisica Nucleare, Sezione di Cagliari, 09042 Monserrato (Cagliari), Italy,
\ensuremath{^{c}}University of California Irvine, Irvine, CA 92697, USA,
\ensuremath{^{d}}Institute of Physics, Academy of Sciences of the Czech Republic, 182~21, Czech Republic,
\ensuremath{^{e}}CERN, CH-1211 Geneva, Switzerland,
\ensuremath{^{f}}Cornell University, Ithaca, NY 14853, USA,
\ensuremath{^{g}}University of Cyprus, Nicosia CY-1678, Cyprus,
\ensuremath{^{h}}Office of Science, U.S. Department of Energy, Washington, DC 20585, USA,
\ensuremath{^{i}}University College Dublin, Dublin 4, Ireland,
\ensuremath{^{j}}ETH, 8092 Z\"{u}rich, Switzerland,
\ensuremath{^{k}}University of Fukui, Fukui City, Fukui Prefecture, Japan 910-0017,
\ensuremath{^{l}}Universidad Iberoamericana, Lomas de Santa Fe, M\'{e}xico, C.P. 01219, Distrito Federal,
\ensuremath{^{m}}University of Iowa, Iowa City, IA 52242, USA,
\ensuremath{^{n}}Kinki University, Higashi-Osaka City, Japan 577-8502,
\ensuremath{^{o}}Kansas State University, Manhattan, KS 66506, USA,
\ensuremath{^{p}}Brookhaven National Laboratory, Upton, NY 11973, USA,
\ensuremath{^{q}}Istituto Nazionale di Fisica Nucleare, Sezione di Lecce, Via Arnesano, I-73100 Lecce, Italy,
\ensuremath{^{r}}Queen Mary, University of London, London, E1 4NS, United Kingdom,
\ensuremath{^{s}}University of Melbourne, Victoria 3010, Australia,
\ensuremath{^{t}}Muons, Inc., Batavia, IL 60510, USA,
\ensuremath{^{u}}Nagasaki Institute of Applied Science, Nagasaki 851-0193, Japan,
\ensuremath{^{v}}National Research Nuclear University, Moscow 115409, Russia,
\ensuremath{^{w}}Northwestern University, Evanston, IL 60208, USA,
\ensuremath{^{x}}University of Notre Dame, Notre Dame, IN 46556, USA,
\ensuremath{^{y}}Universidad de Oviedo, E-33007 Oviedo, Spain,
\ensuremath{^{z}}CNRS-IN2P3, Paris, F-75205 France,
\ensuremath{^{aa}}Universidad Tecnica Federico Santa Maria, 110v Valparaiso, Chile,
\ensuremath{^{bb}}Sejong University, Seoul 143-747, Korea,
\ensuremath{^{cc}}The University of Jordan, Amman 11942, Jordan,
\ensuremath{^{dd}}Universite catholique de Louvain, 1348 Louvain-La-Neuve, Belgium,
\ensuremath{^{ee}}University of Z\"{u}rich, 8006 Z\"{u}rich, Switzerland,
\ensuremath{^{ff}}Massachusetts General Hospital, Boston, MA 02114 USA,
\ensuremath{^{gg}}Harvard Medical School, Boston, MA 02114 USA,
\ensuremath{^{hh}}Hampton University, Hampton, VA 23668, USA,
\ensuremath{^{ii}}Los Alamos National Laboratory, Los Alamos, NM 87544, USA,
\ensuremath{^{jj}}Universit\`{a} degli Studi di Napoli Federico I, I-80138 Napoli, Italy
}
\noaffiliation
\date{\today}

\begin{abstract}
A search for a Higgs boson 
with suppressed couplings to fermions, $h_f$, 
assumed to be the neutral, lower-mass partner of 
the Higgs boson discovered at the Large Hadron Collider, is reported. 
Such a Higgs boson could exist 
in extensions of the standard model with two Higgs doublets,
and could be produced via 
$p\bar{p} \to H^\pm h_f \to W^* h_f h_f \to 4\gamma + X$,
where $H^\pm$ is a charged Higgs boson.
This analysis uses all events with at least three photons in the final state
from proton-antiproton collisions 
at a center-of-mass energy of 1.96~TeV 
collected by the Collider Detector at Fermilab,
corresponding to an integrated luminosity of 9.2~${\rm fb}^{-1}$. 
No evidence of a signal is observed in the data.
Values of Higgs-boson masses
between 10 and 100 \mgev\ 
are excluded at 95\% Bayesian credibility.

\end{abstract}


\maketitle
\clearpage
%
In the standard model (SM) of particle physics, 
the masses of elementary particles are 
generated by the spontaneous breaking of the electroweak 
gauge symmetry~\cite{Higgs},
which predicts the existence of the Higgs boson.
In 2012, 
the ATLAS and CMS experiments at CERN's Large Hadron Collider (LHC)
discovered a scalar boson with mass of approximately 125~\mgev\
and properties consistent with those expected 
for the SM Higgs boson~\cite{LHCDiscovery, LHCFermion}.
Some evidence for such a boson had also been presented 
by the Tevatron experiments~\cite{TeVEvidence}.
The detailed phenomenology of the Higgs boson is, however, 
yet to be investigated.
The possibility that the recently observed Higgs boson is part of
an extended Higgs sector is attractive because
it would address some relevant open questions about the SM
and 
it is not ruled out experimentally.

A minimal extension,
the ``two-Higgs-doublet model'' (2HDM)~\cite{2HDM},
assumes
two doublets of Higgs fields.
The resulting particle spectrum
for the {\it CP}-conserving case
consists of 
three electrically neutral Higgs bosons, $h^0$, $H^0$ and $A^{0}$,
and 
two charged Higgs-bosons, $H^{+}$, $H^{-}$, 
where $h^0$ is less massive than $H^0$.
The acronym {\it CP} represents the combined operations of charge-conjugation
and parity transformation.
An important parameter for predictions from the model
is the ratio 
$\tan\beta$
of the two vacuum-expectation values
for the neutral components of the two Higgs doublets.
Assuming that the boson discovered recently at the LHC is the $h^0$, 
searches for additional, more-massive neutral Higgs bosons 
were performed~\cite{2HDM-CDF, 2HDM-LHC}, 
yielding exclusion limits on production cross sections.

In this Letter, we consider an alternative case in which
the newly-discovered boson corresponds to the high-mass $H^0$
and the lower-mass $h^0$ is yet to be observed.
This scenario is poorly constrained experimentally
if $\tan\beta$ is large and 
$h^0$ has suppressed couplings to fermions at leading order.
The $h^0$
is referred to as the fermiophobic Higgs boson ($h_f$).
Searches performed
at various experiments~\cite{LEP, TevatronRunIILimits, LHC-FP}
have set
lower bounds of its mass, \hmass, at 100--150~\mgev.
These mass limits, however, were obtained assuming 
simplified models in which
the couplings between the $h_f$ and electroweak-gauge bosons 
are
of the same strength as those in the SM,
which is not necessarily true in the 2HDM, 
as they may be strongly suppressed 
when $\tan\beta$ is large~\cite{HiggsHuntersGuide}, 
by a factor of approximately $10^{-2}$ when $\tan\beta = 10$, for example.
A low-mass
$h_f$ ($\hmass \lesssim 100$~\mgev), therefore, could have
eluded the previous searches if $\tan\beta$ is large.
To fill
this gap in exploring the Higgs sector,
we focus on the process
$
q\bar{q}'\rightarrow W^{\ast} \rightarrow h_f H^{\pm}, 
$
followed by the decay
$
H^{\pm} \rightarrow h_f W^*,
$
where $q$ and $\bar{q}'$ are quarks and antiquarks in the colliding
protons and antiprotons taking part in the hard interaction, 
and $W^*$ represents a virtual $W$ boson.
This process, involving 
$H^\pm$, 
has enhanced production rates for large $\tan\beta$~\cite{FP}.
By assuming no couplings to fermions, 
the branching fraction (\BF) of $h_f$ decays to two photons, 
$
h_f\rightarrow \gamma\gamma
$
, 
is near 100\% for $\hmass\lesssim 95$~\mgev~\cite{FP,FPDecays}.
The production of two $h_f$ particles could result 
in a distinctive multiphoton topology with small background rates.
The couplings of the $H^0$ to SM particles in this scenario are similar to 
those of the SM Higgs boson~\cite{FP} and 
we perform the analysis assuming that its mass, $m_{H^0}$, is 125~\mgev.
We also assume 
the $A^{0}$ mass, $m_{A^0}$, to be 350~\mgev,
large enough so as not to contribute to $H^{\pm}$ decays --
the specific choice of $m_{A^0}$ has little effect on the 
final result,
and
we take $\tan\beta=10$.

This analysis is based on the entire data set
of proton-antiproton collisions at a center-of-mass energy of 1.96 TeV
collected with the Collider Detector at Fermilab (CDF II)
between February 2002 and September 2011,
corresponding to an integrated luminosity of 9.2 fb${}^{-1}$.
We select events with multiple photon candidates 
by applying criteria optimized 
for achieving the best sensitivity.
We compare the observed event yields with background expectations,
which are evaluated using a combination of Monte Carlo (MC)
simulation and experimental data.
A challenge is to estimate the contribution from 
background events containing clusters of particles (jets) 
misidentified as photons.

CDF II is a general-purpose detector
consisting of tracking devices in a 1.4 T axial magnetic field, 
surrounded by calorimeters with a projective-tower geometry, 
and muon detectors surrounding the calorimeters.
Gas proportional wire chambers with cathode strips
(shower-maximum strip detectors) are
located
at a depth approximately corresponding to the maximum development of
typical electromagnetic (EM) showers
to measure precisely their centroid position and shape
in the plane transverse to the shower development.
Detailed descriptions of the CDF II detector are in Ref.~\cite{CDF}.

The initial data sample is obtained using a real-time 
event-selection system (trigger) that requires either
two EM-energy clusters in the calorimeter,
each with $\et \equiv E\sin\theta > 12$ GeV, or
three clusters, each with $\et > 10$ GeV, 
where $E$ is the cluster energy measured with the calorimeter,
$\theta$ is the polar angle, 
and $\et$ is the transverse energy~\cite{CDFCoordinate}.
In the analysis,
we select events
with at least three EM energy clusters with $\et>15$ GeV
in the central detector 
(pseudorapidity magnitude $|\eta|<$ 1.1)~\cite{CDFCoordinate}.
The photons
are also required to be isolated:
additional calorimeter \et\
in a cone of angular radius 
$R=\sqrt{(\Delta\eta)^2+(\Delta\phi)^2}=0.4$~\cite{CDFCoordinate}
around the photon candidate must be less than 2 GeV, 
and the scalar sum of transverse momenta of charged particles
in the same cone must be less than
2 \pgev.
We then apply photon-identification criteria based on 
the EM-shower profile, which must be consistent with the expectation 
for an isolated photon.


We estimate 
the reconstruction efficiency for signal events
as a function of \hmass,
from \mhMin\ to \mhMax~\mgev\ with a typical step size of 5 \mgev,
and of the $H^\pm$ mass, \Hmass,
from \mHMin\ to \mHMax~\mgev\ with a typical step size of 10 \mgev,
using \pythia\ (version 6.4) MC simulation~\cite{PYTHIA}.
The generated events are passed through the full detector simulation
based on \geant~\cite{GEANT}.
The simulation of the EM response of the detector
is calibrated by matching the observed energies in samples 
of $Z \to e^+e^-$ events in the data and the MC simulation~\cite{CDFPhotonID}.
The efficiencies,
before applying any further selection criteria to increase the search 
sensitivity,
are within 1--10\%, depending on \hmass\ and \Hmass.

\def \bgdir {./bg}
\def \bestHT {90~GeV}
Direct triphoton production is a major source of background events.
We predict the kinematic distributions
from simulated data generated with
\mg\ (version 5) interfaced with \me~\cite{MGME}
and combined with parton showering from \pythia.
\mgme\ provides 
direct triphoton production with up to two additional jets.
The generated events
are passed through the full detector simulation
and 
we apply the same photon selection as that used for data.

Another source of background is the production of events with
jets misidentified as photons.
For estimating this contribution,
we introduce a loose photon selection
by choosing a subset of the selection criteria.
In a sample of three-photon candidates selected with the loose selection,
there are eight possible combinations of $E_T$-ordered 
photons and EM-like jets,
$\gamma \gamma \gamma$,
$\gamma \gamma j$, $\cdots$,
where $j$ represents an EM-like jet.
The numbers of these events are unknown and we express them by a vector
$
\Vntrue
$
of event counts
($n^*_{\gamma\gamma\gamma}$, $n^*_{\gamma\gamma j}$, $\cdots$).
By applying the full set of criteria for the photon selection,
we
categorize the events in eight classes depending on 
whether each of the photon candidates
in a given event passes ($p$) or fails ($f$)
the full photon selection ($n_{ppp}$, $n_{ppf}$, $\cdots$),
denoted
by $\Vnobs$.
The components of $\Vntrue$
are obtained by solving eight linear equations
$\Vnobs = \E \Vntrue$, 
where \E\ is an $8\times8$ matrix, the elements of which
are calculated from the probability for a genuine photon or jet 
that meets the loose selection to also meet the full photon selection.
Once $\Vntrue$ is obtained by inverting the matrix \E, 
we estimate the misidentified-jet contribution
to $n_{ppp}$ using \bm{E} and 
the calculated elements of $\Vntrue$ except $n^*_{\gamma\gamma\gamma}$.
Statistical uncertainties are propagated to $\Vntrue$.
We estimate
the probability for misidentifying jets as photons
as a function of \et\ 
using isolated jets 
in data samples collected with inclusive jet triggers.
We correct for contributions of genuine photons to the objects 
passing the photon selection in the jet samples, 
which is approximately 70\%, 
based on the differences in the expected distributions of 
isolation and shower shape 
between the misidentified jets and genuine photons.
The misidentification probability 
varies from a few percent to 25\% depending on the \et.

A third source of background events arises
from electroweak processes containing
$Z(\to ee)\gamma      $,
$W(\to e\nu)\gamma    $,
$Z(\to\tau\tau)\gamma $,
or
$W(\to\tau\nu)\gamma$
decays
with
additional misidentified jets or other photon-like particles
that result in the $\gamma\gamma\gamma$ signature.
We investigate these processes using MC simulation and 
calibrate the rates 
with experimental data.

The total expected number of background events at this stage is $10.3 \pm 0.2$,
where the uncertainty is statistical.
We observe 10 events in the data, 
which is consistent with the background expectation.
None of the observed events contains four or more photons.

\begin{figure}[tb]
\centering
\TwoGSizeNote{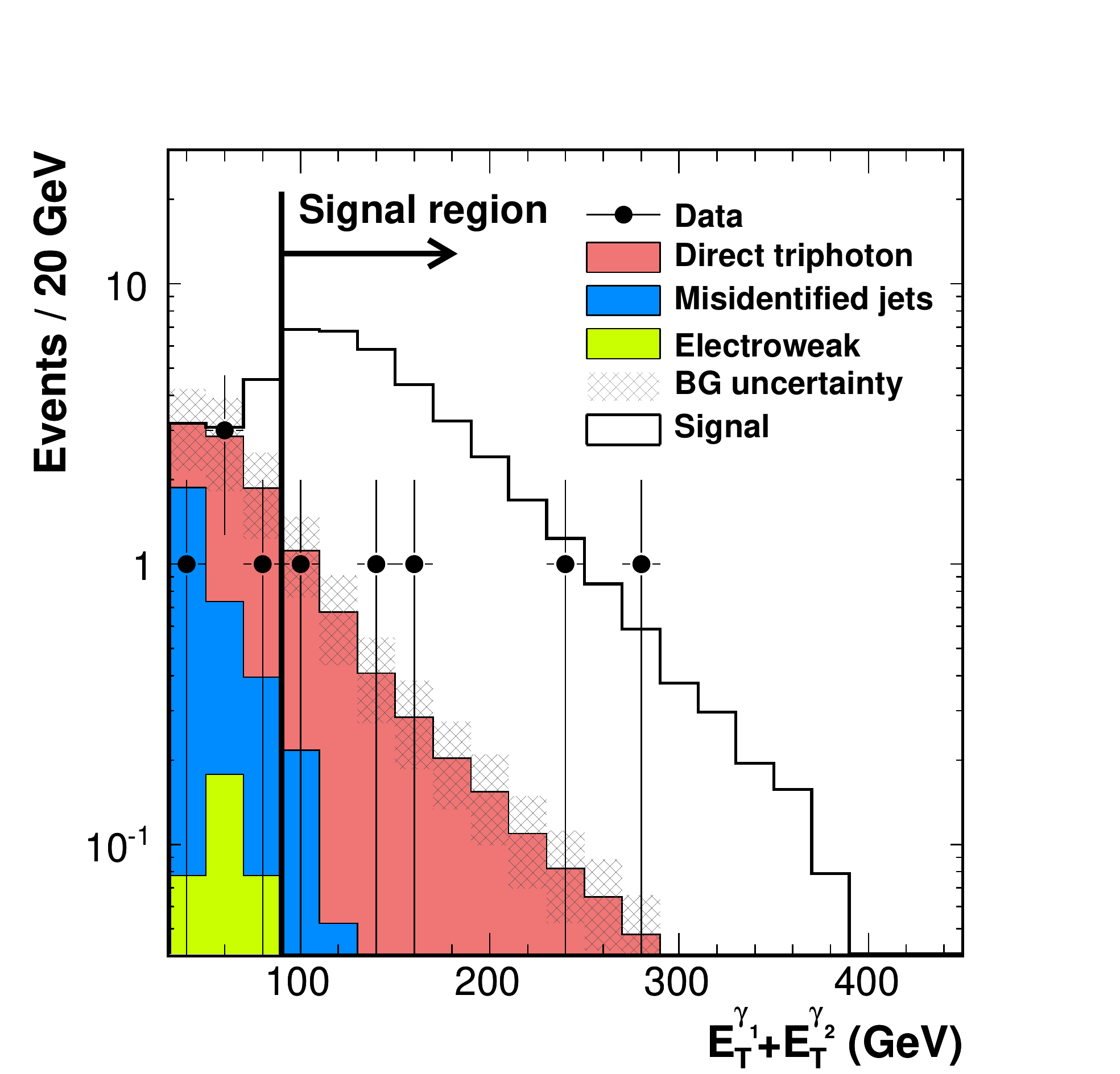}
\caption{Distribution of \etsum\ 
in events containing three or more photons for data, 
SM background prediction, 
and hypothetical signal
for \exampleMasses.
}
\label{fig:Et1Et2}
\end{figure}

In order to further improve the search sensitivity, 
we apply an additional criterion
on the summed \et\ of the two highest-\et\ photons, \etsum.
To quantify the search sensitivity,
we calculate Bayesian~\cite{Bayes} expected limits
on the product of the \xsec\ and the branching fraction
\begin{displaymath}
\sigma(p\bar{p}\rightarrow h_f H^{\pm})
\times \BF(H^{\pm}\rightarrow h_fW^*) 
\times \left[\BF(h_{f}\rightarrow \gamma\gamma)\right]^{2} \msp, 
\end{displaymath}
with respect to theoretical predictions
by integrating posterior probability density functions
based on predicted number of background events.
We assume a uniform prior probability densityfor the signal rate.
The theoretical \xsec s at leading order are computed using \pythia\ 
with an enhancement factor of 1.4
to approximate higher-order contributions,
based on theoretical estimation 
and cross-section measurements of known processes~\cite{dXSec}.
The branching fractions are calculated with
the
{\sc 2hdmc} program (version 1.6.5)~\cite{2HDMC}.
The expected limit is
the median in a large set of simulated experiments 
based on the Poisson fluctuation of the background events.
We choose $\etsum >$ \bestHT\ as the final requirement
because it provides the best expected limit.
Figure~\ref{fig:Et1Et2} shows the predicted and observed distributions of
\etsum\
and includes
the requirement defining the signal region.
We compare the background distribution
and the expected signal distribution
for \exampleMasses.

The main systematic uncertainty on the signal efficiency
comes from that on the estimation of the identification efficiency 
for three photons,
which is 8\% of the total efficiency based on studies comparing $Z \to e^+e^-$ 
in data and simulation~\cite{CDFPhotonID}
by assuming full correlation among three photons.
Other sources of systematic uncertainties include those on
the parton momentum distributions in the colliding hadrons,
the initial-
and 
final-state radiation of a gluon,
and
the renormalization scale,
which are each found to contribute less than 3\% of the total efficiency.

We compare the \mgme\ \xsec\
with \mcfm~\cite{NLO} calculations that take into account different
higher-order contributions
and
take the resulting difference of
\DTPSysXSec\ as a systematic uncertainty on the yield of direct triphoton
events.
The systematic uncertainty from
the renormalization scale, 
that from the initial- and final-state radiation, 
and that from the luminosity measurement~\cite{LuminositySys}
range from 0.16 to 0.21 events.
We estimate
the total systematic uncertainty on the 
expected yield of events with 
misidentified jets
to be 0.17 events,
which includes the contribution
from the measurement of the misidentified-jet probability
and
that from the possible difference of the probabilities
between jets originating from quarks and gluons.
The dominant uncertainty on the electroweak contribution 
originates from the limited size of the simulated event samples
used 
to estimate the small probability to find an extra photon-like particle in
the $W(\to e\nu)\gamma$ events.

Table~\ref{tab:final_bg} shows the expected number of background events
and
the number of events found in data after the final selection.
\begin{table}[tb]
\caption{Expected number of background events compared to the 
observed number of events after the final event selection.
The first contribution to the uncertainty is statistical
and 
the second is systematic.
}
\label{tab:final_bg}
\renewcommand{\arraystretch}{1.2}
\bcenter
\begin{tabular}{lrcccc}
\hline\hline
&\mc{5}{c}{Events in signal region} \\
&\mc{5}{c}{($E_{T}^{\gamma_1}+E_T^{\gamma_2}>90$ GeV)} \\
\hline
Direct triphoton  &\makebox[8mm]{2.60}
&\makebox[8mm]{$\pm$}
&\makebox[8mm]{0.04}
&\makebox[8mm]{$\pm$}
&\makebox[8mm]{0.93}
\\
Misidentified jets \hspace{8mm} &\makebox[8mm]{0.32}
&\makebox[8mm]{$\pm$}
&\makebox[8mm]{0.07}
&\makebox[8mm]{$\pm$}
&\makebox[8mm]{0.17}
\\
Electroweak  &\makebox[8mm]{0.04}
&\makebox[8mm]{$\pm$}
&\makebox[8mm]{0.01}
&\makebox[8mm]{$\pm$}
&\makebox[8mm]{0.03}
\\
\hline
Total &\makebox[8mm]{2.96}
&\makebox[8mm]{$\pm$}
&\makebox[8mm]{0.08}
&\makebox[8mm]{$\pm$}
&\makebox[8mm]{0.94}
\\
\hline
Data &5\hspace{15pt} \\
\hline\hline
\end{tabular}
\ecenter
\renewcommand{\arraystretch}{1}

\end{table}
We find 5 candidate events in data, 
which is consistent with the expected number of background events.


We check the background predictions using
background-rich control samples.
In events
containing one lower-quality photon candidate that
passes the loose selection but fails the full selection,
the predicted and observed numbers of events are 
$372 \pm 68$ and $370$, respectively.
In events with
$\etsum <$ \bestHT,
$6.6 \pm 1.7$ events are predicted and 5 events observed.
The observed
agreement supports the reliability of the background estimation.

\begin{figure}[tb]
\centering
\TwoGSizeNote{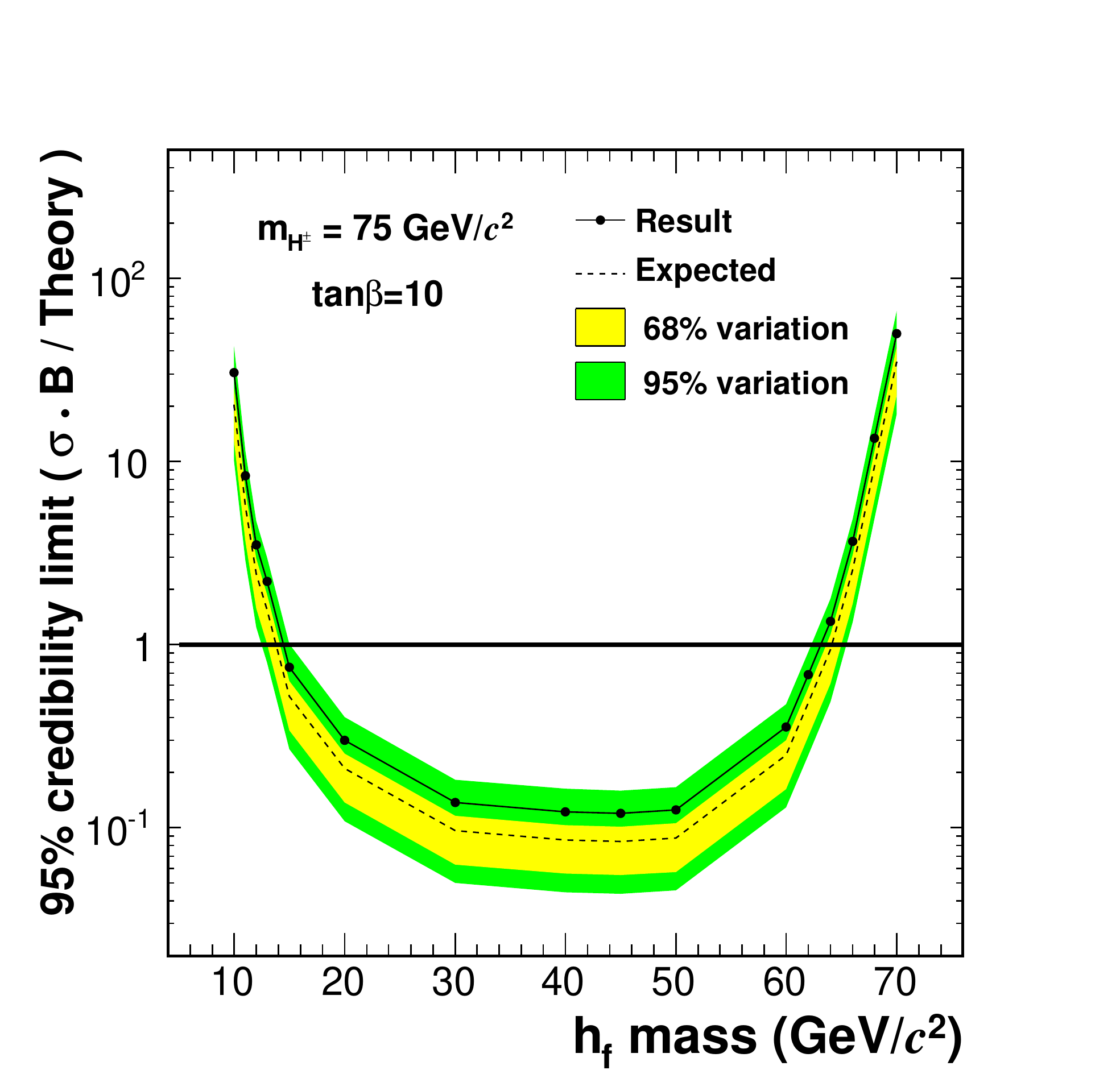}
\caption{
Upper limit at 95\% \CL\ on the \xsechyph\ ratio with respect to 
theory predictions, 
calculated for 
the final selection, including the
$\etsum >$ \bestHT\ requirement.
The solid line is the obtained limit, 
the dashed line is the expected limit, and the shaded regions cover
the 68\% and 95\% of possible variations of expected limit values
based on the Poisson statistics of the expected number of background events.
}
\label{fig:exLim95b}
\end{figure}

\begin{figure}
\centering
\TwoGSizeNote{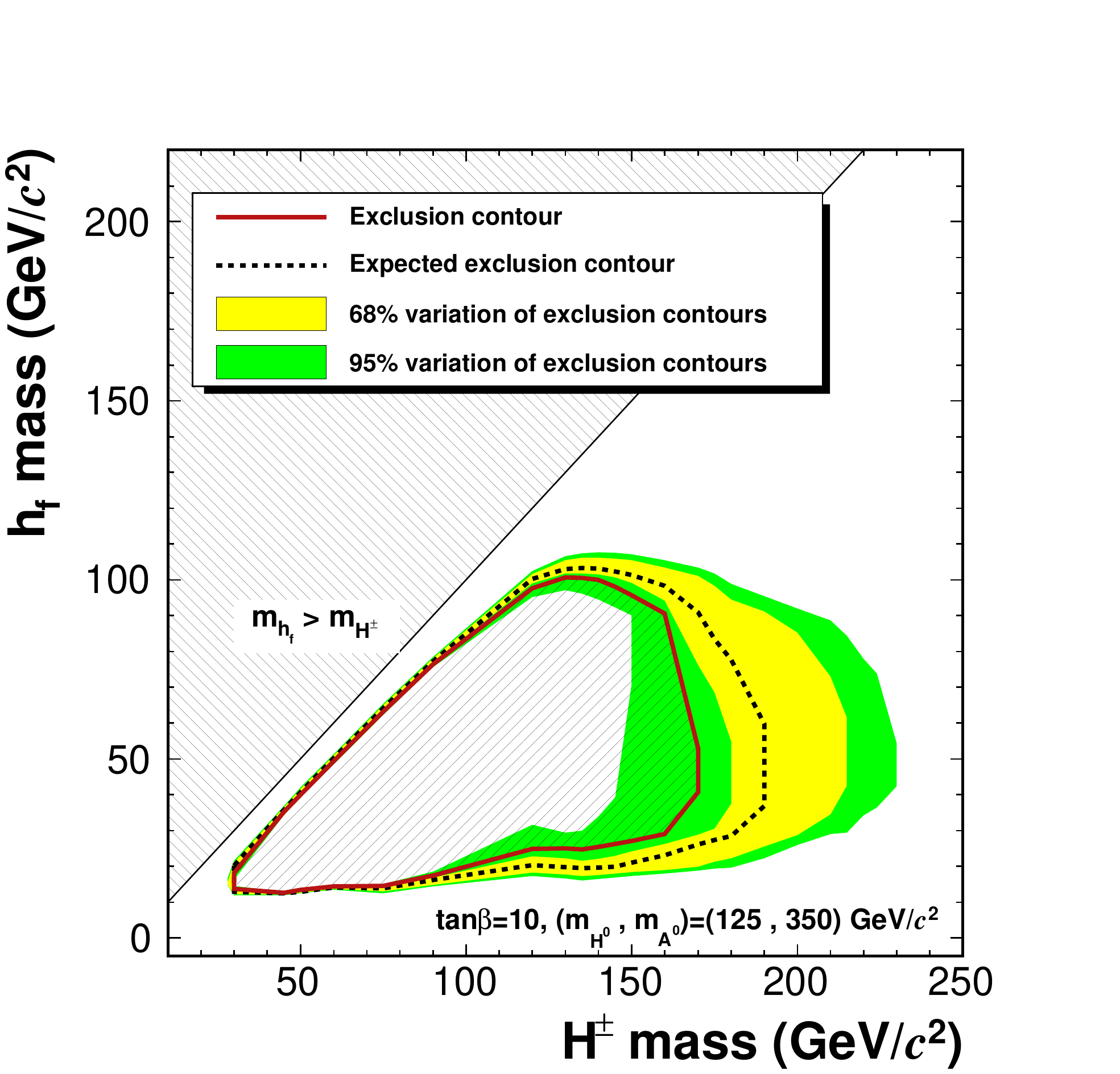}
\caption{
Excluded mass region at a 95\% \CL, 
calculated for 
the final selection.
The solid curve is the contour enclosing the exclusion region,
the dashed line encloses the median expected exclusion region,
and the shaded regions cover
the 68\% and 95\% of possible variations of expected contours
based on the Poisson statistics of the expected number of background events.
}
\label{fig:exLim95Region}
\end{figure}

We perform a Bayesian limit calculation restricted to events observed
in the signal region, $\etsum>$ \bestHT,
as a function of \hmass, ranging from \mhMin\ to \mhMax~\mgev,
and
\Hmass, ranging from \mHMin\ to \mHMax~\mgev.
We include
systematic uncertainties 
due to
the signal efficiency, 
the predicted number of background events, and the luminosity, 
as well as the theoretical uncertainty of 20\% 
on the \xsec\ of Higgs boson production~\cite{dXSec}.
Figure~\ref{fig:exLim95b}
shows the expected and the observed \xsec\ limits 
at 95\%~\CL\ for a particular choice of \hmass\ and \Hmass,
with possible variations of the expected limits obtained by assuming
68\% or 95\% of Poisson fluctuations of 
the number of background events.
From Fig.~\ref{fig:exLim95b}, 
the \hmass\ region betwen 14 and 62~\mgev\ is excluded 
for $\Hmass=75$~\mgev.
Connecting the boundary regions 
of the excluded \hmass\ region for various values of
\Hmass\ in the \hmass\ vs. \Hmass\ plane,
we form contours of the excluded mass regions and 
present them
in Fig.~\ref{fig:exLim95Region}.
The region of parameters given by \hmass\ between 10 and 100 \mgev\
and \Hmass\ between 30 and 170 \mgev\ is excluded.
The result does not change significantly if we repeat the analysis 
by assuming
$\tan\beta = 30$, while the excluded region shrinks by approximately 20 \mgev\
for both of \hmass\ and \Hmass\ for $\tan\beta = 3$.

In conclusion, 
we report on a search for the fermiophobic Higgs boson
in the two-Higgs-doublet model
using events with at least three photons in the
final state, resulting from the hypothetical process
$
p\overline{p} \to h_f H^\pm
$
followed by $H^\pm \to h_f W^*$ and $h_f \to \gamma\gamma$.
The observed number of signal candidate events in data
is consistent with 
the expected number of background events.
We calculate
the upper limit on the product of the \xsec\ and the branching
fraction at 95\% Bayesian credibility
for \hmass\ values ranging
from \mhMin\ to \mhMax~\mgev\ 
and for \Hmass\ values ranging from \mHMin\ to \mHMax~\mgev,
and then translate these limits
into an excluded region in the
\hmass\ vs. \Hmass\ plane,
shown in Fig.~\ref{fig:exLim95Region}.
The region of parameters given by \hmass\ between 10 and 100 \mgev\
and \Hmass\ between 30 and 170 \mgev\ is 
excluded for $\tan\beta=10$.
This is
the first search
for a fermiophobic neutral Higgs boson 
with mass smaller than the boson discovered at the LHC
in the two-Higgs-doublet model.

\begin{acknowledgments}
We thank the Fermilab staff and the technical staffs of the
participating institutions for their vital contributions. This work
was supported by the U.S. Department of Energy and National Science
Foundation; the Italian Istituto Nazionale di Fisica Nucleare; the
Ministry of Education, Culture, Sports, Science and Technology of
Japan; the Natural Sciences and Engineering Research Council of
Canada; the National Science Council of the Republic of China; the
Swiss National Science Foundation; the A.P. Sloan Foundation; the
Bundesministerium f\"ur Bildung und Forschung, Germany; the Korean
World Class University Program, the National Research Foundation of
Korea; the Science and Technology Facilities Council and the Royal
Society, United Kingdom; the Russian Foundation for Basic Research;
the Ministerio de Ciencia e Innovaci\'{o}n, and Programa
Consolider-Ingenio 2010, Spain; the Slovak R\&D Agency; the Academy
of Finland; the Australian Research Council (ARC); and the EU community
Marie Curie Fellowship Contract No. 302103.

\end{acknowledgments}



\begin{thebibliography}{1}
\bibitem{CDFCoordinate}
CDF uses a cylindrical 
coordinate system with $+z$ in the proton beam direction,
$\theta$ and $\phi$ are the polar and azimuthal angles, respectively,
and $\eta$ is the pseudorapidity defined by $\eta \equiv -\ln\tan(\theta/2)$.
\end{thebibliography}
\end{document}